\documentclass{svproc}
\pdfoutput=1
\usepackage[normalem]{ulem}
\usepackage[utf8]{inputenc}
\usepackage{graphicx,bm}
\usepackage{slashed,verbatim}
\usepackage{amssymb,graphicx,epstopdf}

 \usepackage[colorlinks=true,linktocpage=true,linkcolor=blue,citecolor=blue]{hyperref}
\usepackage{slashed,subfigure}
\usepackage{caption}
\usepackage{amsmath,amssymb,graphicx,xcolor,mathtools}
\usepackage{epstopdf,dcolumn}
\usepackage{soul}

\allowdisplaybreaks
\usepackage[normalem]{ulem}
\usepackage{cancel}
\usepackage{xcolor}
\setstcolor{red}
\usepackage{url,graphicx,multicol,footmisc,verbatim}

\def\be {\begin{equation}}
\def\ee {\end{equation}}
\def\bea {\begin{eqnarray}}
\def\eea {\end{eqnarray}}
\def\bc {\begin{center}}
\def\ec {\end{center}}
\def\nn {\nonumber}

\def\mn {\mu\nu}

\def\({\left(}
\def\){\right)}
\def\[{\left[}
\def\]{\right]}

\begin{document}
\mainmatter              % start of a contribution
\title{General structure of a gauge boson propagator and pressure of deconfined QCD matter in a weakly magnetized medium}
\titlerunning{General structure of a gauge boson propagator and pressure of}  % abbreviated title (for running head)
%                                     also used for the TOC unless
%                                     \toctitle is used
%
\author{Bithika Karmakar\inst{1}
\and 
Aritra Bandyopadhyay\inst{2}
\and 
Najmul Haque \inst{3}
\and 
Munshi G Mustafa\inst{4}}
% \footnote{Speaker}\thanks{Work done in
%     collaboration with Aritra Bandyopadhyay, Najmul Haque and Munshi G Mustafa}}
%
\authorrunning{Bithika Karmakar et al.} % abbreviated author list (for running head)
%
%%%% list of authors for the TOC (use if author list has to be modified)
%\tocauthor{Bithika Karmakar}
%
% \institute{Saha Institute of Nuclear
%     Physics, HBNI,\\ 1/AF Bidhan Nagar, Kolkata 700064, India\\
% \email{bithika.karmakar@saha.ac.in} \and Departamento de F\'{\i}sica, Universidade Federal de Santa Maria, 
%  	Santa Maria, RS 97105-900, Brazil
%  	
% %  	\email{aritrabanerjee.444@gmail.com} 
% \and School of Physical Sciences, National Institute of Science Education and Research, HBNI,\\  Jatni, Khurda 752050, India}
% %  	\email{nhaque@niser.ac.in}
\institute{Theory Division, Saha Institute of Nuclear Physics, HBNI, 
 	1/AF, Bidhannagar, Kolkata 700064, India ,
\email{bithika.karmakar@saha.ac.in},
\and
Departamento de F\'{\i}sica, Universidade Federal de Santa Maria, 
  	Santa Maria, RS 97105-900, Brazil,
  	\email{aritrabanerjee.444@gmail.com}
  	\and 
  	School of Physical Sciences, National Institute of Science Education and Research, HBNI,  Jatni, Khurda 752050, India,
  	\email{nhaque@niser.ac.in}
  	\and
  	Theory Division, Saha Institute of Nuclear Physics, HBNI, 
 	1/AF, Bidhannagar, Kolkata 700064, India,
 	\email{munshigolam.mustafa@saha.ac.in}}

\maketitle              % typeset the title of the contribution

\begin{abstract}
We have systematically constructed the general structure of the gauge boson self-energy and the effective  propagator in 
presence of a nontrivial background like hot magnetized material medium. Based on this as well as the general structure of fermion propagator in weakly magnetized medium we have calculated pressure of deconfined QCD matter within HTL approximation.  

% \keywords{QCD, Non-central HIC, Weak magnetic field, Thermodynamics}
\end{abstract}
\section{Introduction}
Quark Gluon Plasma is a thermalized color deconfined state of nuclear matter in the 
regime of Quantum Chromo Dynamics (QCD) under extreme conditions such as 
very high temperature and/or density. For the past couple of decades, different high 
energy Heavy-Ion-Collisions (HIC) experiments are under way, \textit{e.g.}, RHIC @ BNL, 
LHC @ CERN and  upcoming  FAIR @ GSI,  to study this novel state of QCD matter. In recent years the noncentral HIC is also being studied, where a very strong magnetic field can be created in 
the direction perpendicular to the reaction plane due to the spectator particles 
that are not participating in the 
collisions~\cite{Shovkovy:2012zn,D'Elia:2012tr,Fukushima:2012vr}. Also some studies have 
showed that the strong magnetic field generated during the 
noncentral HIC is time dependent and  rapidly decreases with 
time~\cite{Bzdak:2012fr,McLerran:2013hla}. At the time of the noncentral HIC, the value of the 
created magnetic field $B$ is very high compared to the temperature $T$ 
($T^2<q_fB$ where $q_f$ is the absolute charge of the quark with flavor $f$)
associated with the system, whereas after few $fm/c$, the magnetic field is shown to decrease to a very low value ($q_fB<T^2$).
In this regime one usually works in the weak magnetic field approximation. 

The presence of an external anisotropic field in the medium calls for the 
appropriate modification of the present theoretical tools to investigate 
various properties of QGP and a numerous activity is in progress.
The  EoS is a generic quantity and of phenomenological
importance for studying the hot and dense QCD matter, QGP, created in HIC.
% In HTLpt the EOS  of QCD \textit{in absence of magnetic field} has systematically been 
% computed  within one-loop(Leading order 
% (LO))~\cite{najmul13,najmul12,najmul11,sylvain2,sylvain1,andersen3,andersen2,andersen1,Haque:2018eph}, 
% two loop (next-LO (NLO))~\cite{Andersen:2002ey,Andersen:2003zk,Haque:2012my,najmul2qns} 
% and three loop (next-to-NLO 
% (NNLO))~\cite{3loopglue1,3loopglue2,3loopqcd1,3loopqcd2,3loopqcd3,najmul3,Haque:2014rua} 
% at finite temperature and chemical potential.

\section{General structure of gauge boson propagator}
Finite temperature breaks the boost symmetry of a system whereas magnetic field or anisotropy breaks the rotational symmetry. We consider the momentum of gluon as $P_\mu=(p_0,p_1,0,p_3)$. We work in the rest frame of the heat bath, {\it{i.e,}} $u^\mu=(1,0,0,0)$ and represent the background magnetic field as $n_\mu \equiv \frac{1}{2B} \epsilon_{\mu\nu\rho\lambda}\, u^\nu F^{\rho\lambda} 
 = \frac{1}{B}u^\nu {\tilde F}_{\mu\nu} = (0,0,0,1)$. 
%  We define the following quantities,
%  \bea
%  \bar{u}^\mu &=& u^\mu - \frac{(P\cdot u)P^\mu}{P^2} 
% = u^\mu - \frac{p_0 P^\mu}{P^2},\nn\\
% \bar n^\mu &=& n^\mu-\frac{(P\cdot n)P^\mu}{P^2} -\frac{(\bar u\cdot n)\bar u^\mu}{\bar u^2}.
% \eea
%  
%  We define the following tensors which are linearly independent.
%  \bea
%  B^{\mn}&=&\frac{\bar u^\mu \bar u^\nu}{\bar u^2}\,\,,\,\,Q^{\mn}=\frac{\bar n^\mu \bar n^\nu}{\bar u^2}\,\,,\,\,N^{\mn}=\frac{\bar u^{\mu}\bar n^{\nu}+\bar u^{\nu}\bar n^{\mu}}{\sqrt{\bar u^2}\sqrt{\bar n^2}}\nn\\
%  R^{\mn}&=&\eta^{\mn}-\frac{P^\mu P^\nu}{P^2}-B^{\mn}-Q^{\mn}.
%  \eea
 The general structure of the gauge boson self energy in presence of magnetic field can be written as~\cite{Karmakar:2018aig}
 \bea
\Pi^{\mn} = b B^{\mn} + c R^{\mn} + d Q^{\mn}+ a N^{\mn}, \label{gen_tb}
\eea
where the form factors $b$, $c$, $d$ and $a$ can be calculated as
\bea
b&=&B^{\mn}\Pi_{\mn}\,\,,\,\,c=R^{\mn}\Pi_{\mn}\,\,,\,\,d=Q^{\mn}\Pi_{\mn}\,\,,\,\,a=\frac{1}{2}N^{\mn}\Pi_{\mn}.
\label{ff_g}
\eea
Using Dyson-Schwinger equation, one can write the general structure of gluon propagator as
\bea
\mathcal{D}_{\mn} &=&\frac{\xi P_{\mu}P_{\nu}}{P^4}+\frac{(P^2-d) B_{\mn}}{(P^2-b)(P^2-d)-a^2}+\frac{R_{\mn}}{P^2-c}+\frac{(P^2-b) Q_{\mn}}{(P^2-b)(P^2-d)-a^2}\nn\\
&&+\frac{a N_{\mn}}{(P^2-b)(P^2-d)-a^2}.
\label{gauge_prop}
\eea
% \begin{center}
% \begin{figure}[h]
% \begin{center}
% {\includegraphics[scale=0.43]{pi3_mpi2_10.pdf}}
% 	\caption{Gluon dispersion curves for $\theta_p=\pi/3$ and magnetic field strength 
% 	 $|eB|= m_\pi^2/10$ for $N_f=2$.}
% 	\label{fig:disp_pi3}
% 	\end{center}
% \end{figure}
% \end{center}
It is found from the poles of Eq.~(\ref{gauge_prop}) that the gluon in hot magnetized medium has three dispersive modes which are given as 
\bea
P^2-c &=& 0,\\
(P^2-b)(P^2-d)-a^2&=&(P^2-\omega_n^+)(P^2-\omega_n^-)=0,\label{bd_mode}
\eea
where $\omega_{n^+}=\frac{b+d+\sqrt{\(b-d\)^2+4a^2}}{2}$ and $\omega_{n^-}=\frac{b+d-\sqrt{\(b-d\)^2+4a^2}}{2}$ .
\vspace*{.1in}

We consider small magnetic field approximation and calculate all the quantities up to $\mathcal{O}[\(eB\)^2]$ .
Within this approximation Eq.~(\ref{bd_mode}) becomes
\bea
\(P^2-b\)\(P^2-d\)=0.
\eea

The form factors $b$, $c$ and $d$ are calculated~\cite{Karmakar:2018aig} from Eq.~(\ref{ff_g}) using HTL approximation.

% 
% \section{Quark free energy in the presence of weak magnetic field}
% \label{QF}
% Using the general structure of quark propagator and the form factors in presence of weakly magnetized medium as calculated in~\cite{Das:2017vfh}, one can find the one loop quark free energy as
% \bea
% F_q&=& 
% -\frac{7\pi^2T^4N_cN_f}{180}\left(1+\frac{120}{7}\hat{\mu}^2+\frac{240}{7}
% \hat{\mu}^4\right) \nn\\
% &&\hspace{1cm}- N_c\sum_f\int\frac{d^4P}{(2\pi)^4}~\ln\left[\frac{(A_0+A_s)(A_0-A_s)}{P^4}\right],
% \label{F_qHTL}
% \eea
% where $A_0$ and $A_s$ include quark form factors. $\hat \mu=\mu/2\pi T$ where $\mu$ is the chemical potential. We calculate the one loop quark free energy upto $\mathcal O[g^4]$.
% \section{Gluon free energy  in the presence of weak magnetic field}
% \label{GF}
% We calculate the one loop gluon free energy using the form factors~\cite{Bandyopadhyay:2017cle} in Eq.~(\ref{ff_g}) as
% \bea
% F_g &=& (N_c^2-1)
% \Bigg[\frac{1}{2}\sumintb_{P}~\ln\left(1-\frac{b}{P^2}\right)+\frac{1}{2}\sumintb_{P} ~\ln\left(-P^2 + c\right)+\frac{1}{2}\sumintb_{P} ~\ln\left(-P^2 + d\right)\Bigg].
% \label{free_qed}
% \eea
% Here we note that the Matsubara frequency of gluon is given as $p_0=2n\pi i T$. Thus the momentum of gluon can be hard $(\sim T)$ or soft $(\sim gT)$. We calculate both the hard $(F_g^h)$ and soft $(F_g^s)$ contribution of the one loop gluon free energy.
% \bea
% F_g=F_g^h+F_g^s
% \eea
\section{Free energy and pressure in weak Field approximation}
\label{RFPW}

The total one-loop free energy of deconfined QCD matter in a weakly magnetized  hot medium reads as~\cite{Bandyopadhyay:2017cle}
\bea
F &=& F_q+F_g+F_0 + \Delta {\mathcal E}_0,
\label{total_fef}
\eea
\begin{center}
\begin{figure}[h]
 \begin{center}
 \includegraphics[scale=0.4]{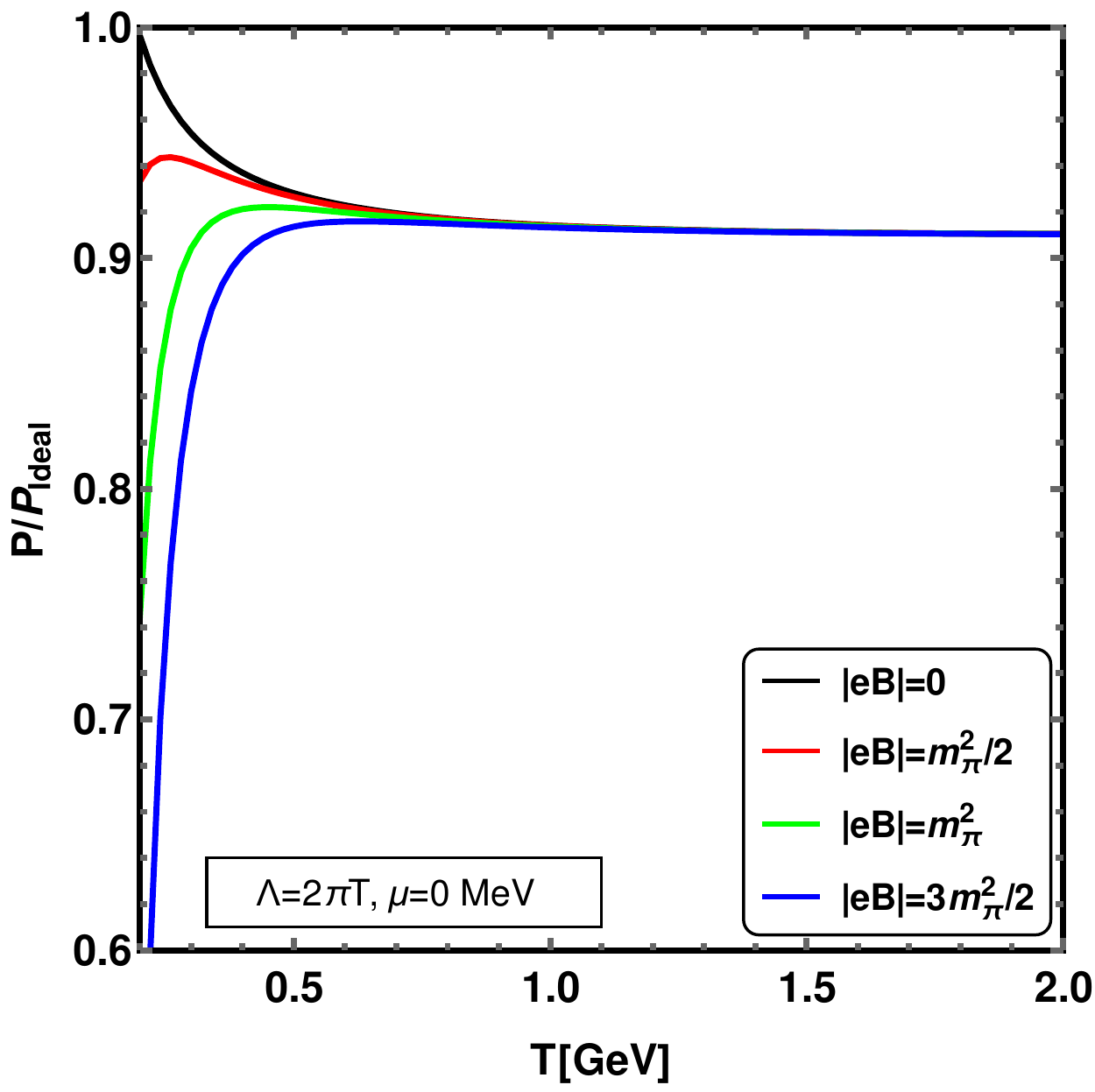}
\includegraphics[scale=0.4]{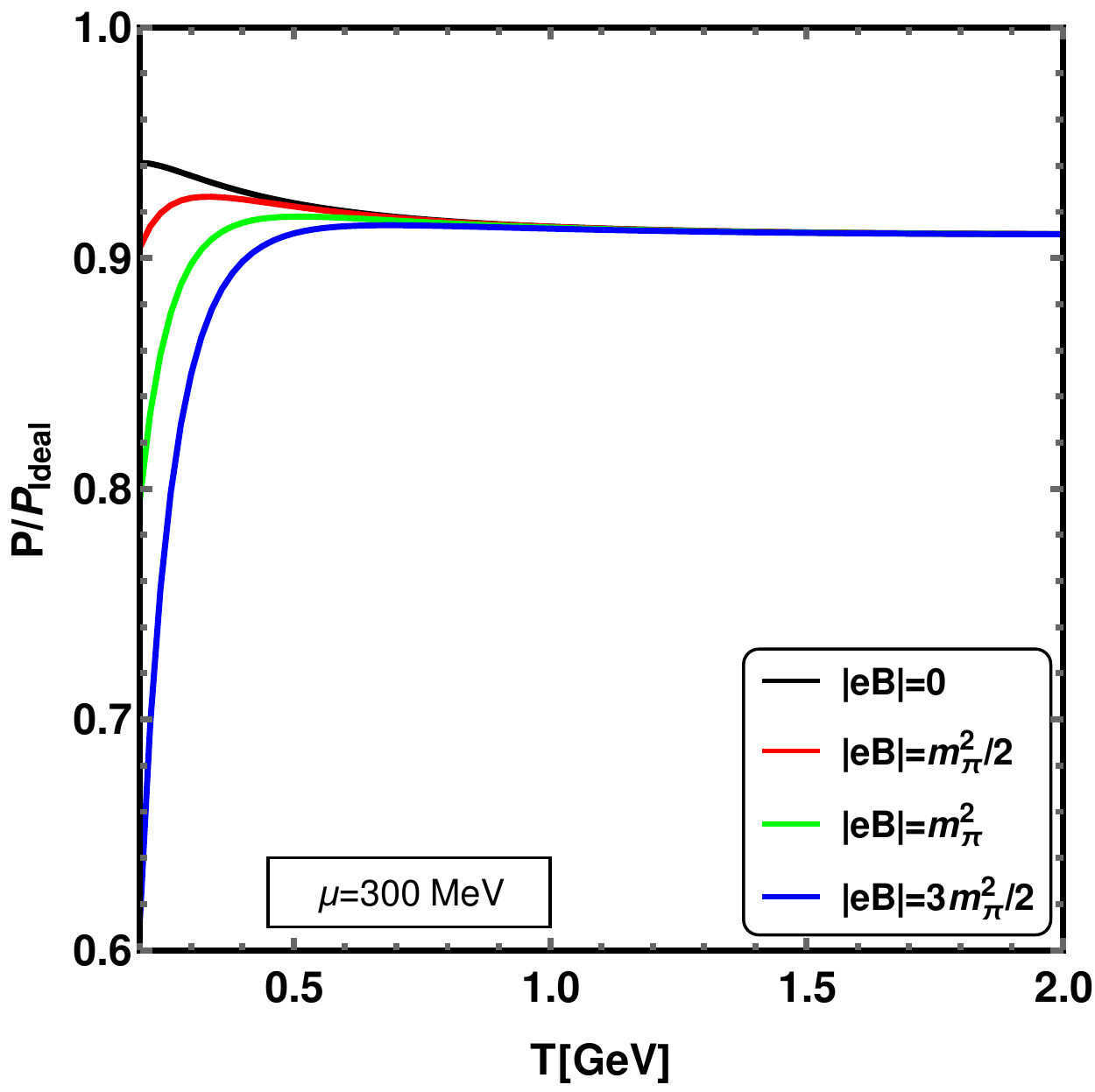} 
 \caption{Variation of the scaled one-loop pressure with temperature for 
$N_f=3$ with $\mu=0$ (left panel) for $\mu=300$ MeV (right panel) in presence of weak 
magnetic field. Renormalization scales are chosen as $\Lambda_g=2\pi T$ for gluon and $\Lambda_q =2\pi \sqrt{T^2+\mu^2/\pi^2}$ for quark.}
  \label{1loop_pressure}
 \end{center}
\end{figure}
\end{center} 
where $F_q$, $F_g$ are quark and gluon free energy in weak magnetized medium which are calculated~\cite{Bandyopadhyay:2017cle} using the form factors corresponding to quark~\cite{Das:2017vfh} and gluon self energy~\cite{Karmakar:2018aig}. $F_0=\frac{1}{2} B^2$ is the tree level contribution due to the constant magnetic field and the $\Delta {\mathcal E}_0$ is the HTL counter term 
given as
\bea
\Delta {\mathcal E}_0=\frac{d_A}{128\pi^2\epsilon}m_D^4, \label{htl_count}
\eea
with $d_A=N_c^2-1$, $N_c$ is the number of color in fundamental representation and $m_D$ is the Debye screening mass in HTL approximation. 
The divergences present in the total free energy are removed by redefining the magnetic field in $F_0$ and by adding counter terms~\cite{Bandyopadhyay:2017cle}.

The pressure of the deconfined QCD matter in weakly magnetized medium is given by
\bea
P(T,\mu,B,\Lambda) = -F(T,\mu,B,\Lambda),
\eea
where $\Lambda$ is the renormalization scale.
% On the other hand the ideal pressure of quark and gluon gas can be calculated as
% \bea
% P_{\text{Ideal}}(T,\mu) = 
% N_cN_f\frac{7\pi^2T^4}{180}\left(1+\frac{120}{7}\hat{\mu}^2+\frac{240}{7}\hat{
% \mu}^4\right) + (N_c^2-1)\frac{\pi^2T^4}{45}.
% \eea

\section{Results}
\vspace*{-.1in}
\begin{center}
\begin{figure}[h]
 \begin{center}
 \includegraphics[scale=0.4]{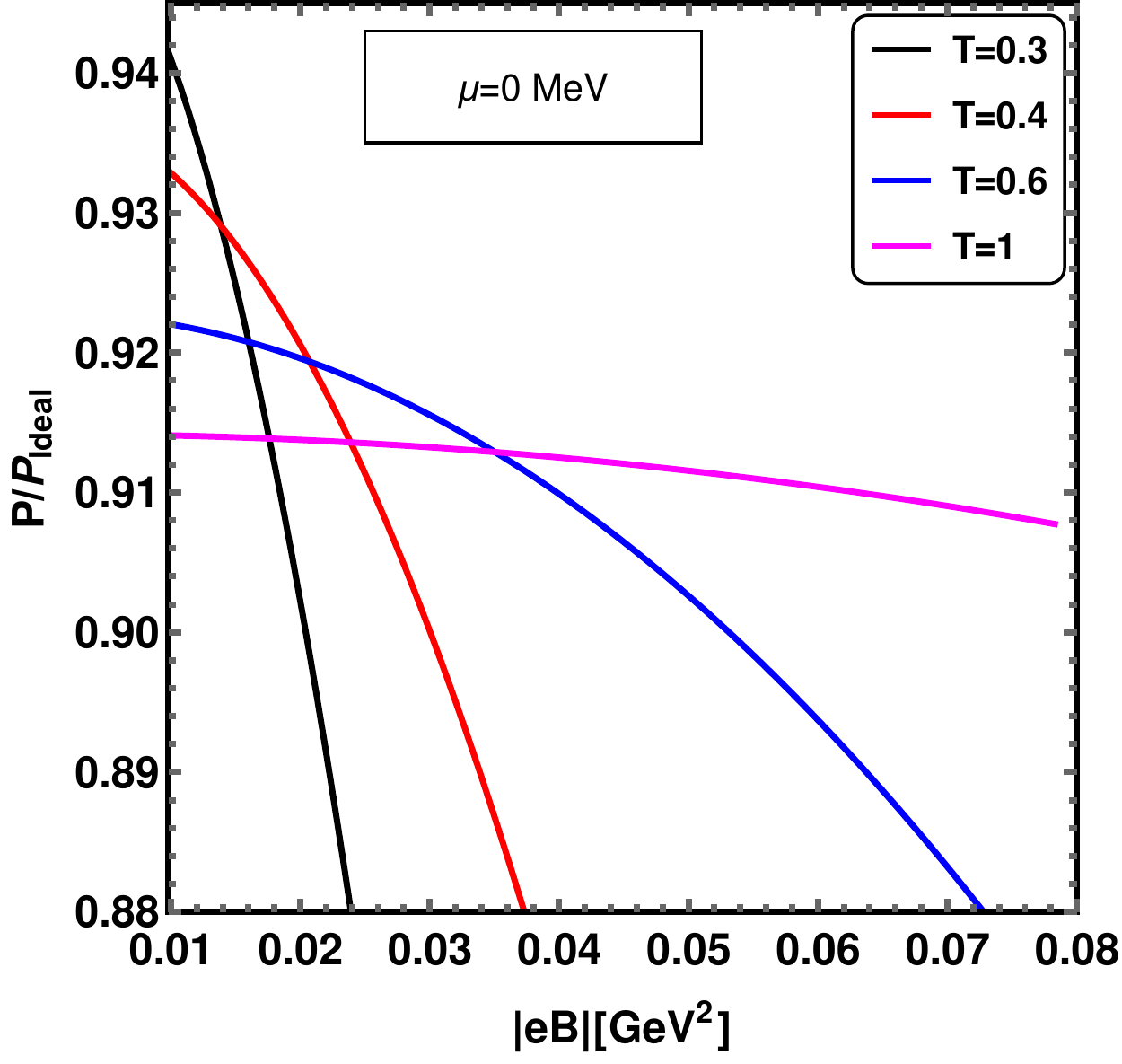} 
\includegraphics[scale=0.4]{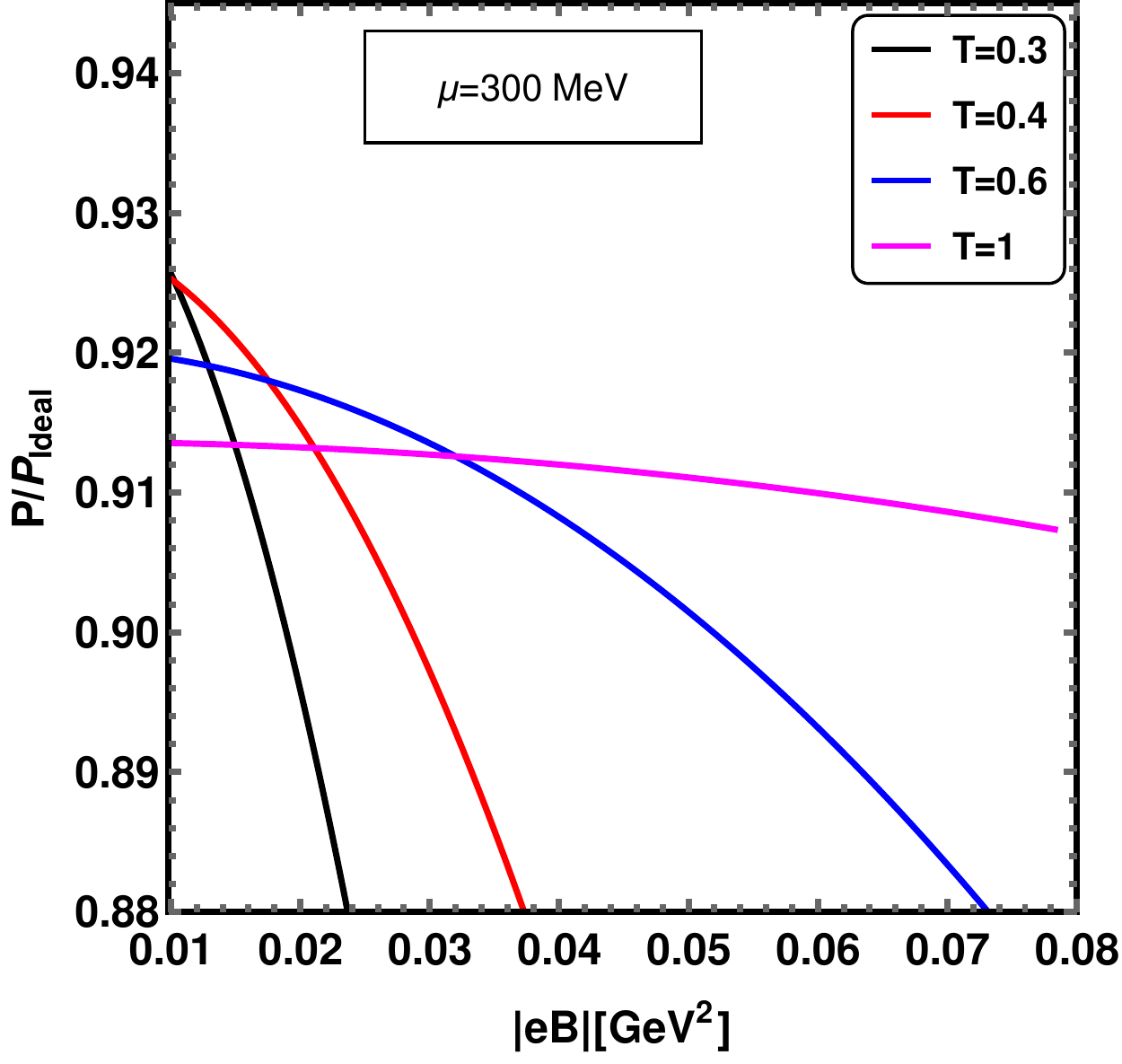} 
 \caption{Variation of the scaled one-loop  pressure with magnetic field for 
$N_f=3$ with $\mu=0$ (left panel) and $\mu=300$ MeV (right panel) for $T=$ (0.3, 0.4, 0.6 and 1) GeV. }
  \label{1loop_pressure_eB}
 \end{center}
\end{figure}
\end{center}
The variation of scaled pressure with temperature is shown in Fig.~\ref{1loop_pressure} for $\mu=0$ and $\mu=300$ MeV. It can be seen from the figure that the magnetic field dependence of the scaled pressure decreases with temperature because temperature is the dominant scale in the weak field approximation $(|eB|<T^2)$ . At high temperature, all the plots for different magnetic field asympotically reach the one-loop HTL pressure. The scaled pressure is plotted with magnetic field strength in Fig.~\ref{1loop_pressure_eB}. We  note from Fig.~\ref{1loop_pressure_eB} that the slopes of the plots decrease with increase of temperature reflecting the reduced magnetic field dependence.

% \begin{center}
% \begin{figure}[h]
%  \begin{center}
%  \includegraphics[scale=0.55]{pressure_varying_Lambda_mu_0_ayala.pdf} 
% \includegraphics[scale=0.55]{pressure_varying_Lambda_mu_300_ayala.pdf} 
%  \caption{Variation of the scaled one-loop pressure with temperature for 
% $N_f=3$ with $\mu=0$ (left panel) and $\mu=300$ MeV (right panel) in presence of a weak 
% magnetic field of strength $eB = m_\pi^2$ for different values of renormalization 
% scale  of gluons, $\Lambda_g=\pi T, \, 2\pi T, \ \textrm{and}\, 4\pi T$ and scale of quarks are given in the legend.}
%   \label{1loop_pressure_vl}
%  \end{center}
% \end{figure}
% \end{center}

\section{Acknowledgment}
BK and MGM were funded by Department of Atomic Energy, India via the 
project TPAES. BK gratefully acknowledges the organizers of XXIII DAE-BRNS High Energy Physics Symposium 2018 for the invitation.

% 
% \begin{comment}
% \section{Remove later}
% With this chapter, the preliminaries are over, and we begin 
% based on symmetry considerations and on pinching estimates, as in
% Sect.~5.2 of this article.
% \end{comment}
%
% ---- Bibliography ----
%

\end{document}